\documentclass[aps,prl,showpacs,twocolumn,amsmath,amssymb]{revtex4-1}
\usepackage{graphicx}
\usepackage{dcolumn}
\usepackage{bm}

\usepackage{multirow}

\usepackage{amsmath}
\usepackage{amssymb}
\usepackage{amsthm}
\usepackage{pinlabel}
\usepackage{natbib}
\usepackage{epstopdf}
\usepackage{color}

\usepackage{braket}

\begin{document}

\title{Multiband, full bandwidth anisotropic Eliashberg theory of interfacial electron-phonon coupling and high-T$_{\rm c}$ superconductivity in FeSe/SrTiO$_3$}
\author{Alex Aperis}\email{alex.aperis@physics.uu.se}
\author{Peter M.~Oppeneer}\email{peter.oppeneer@physics.uu.se}
\affiliation{Department of Physics and Astronomy, Uppsala University, P.\ O.\ Box 516, SE-75120 Uppsala, Sweden}

\vskip 0.4cm
\date{\today}

\begin{abstract}
\noindent 
We examine the impact of interfacial phonons on the superconducting state of FeSe/SrTiO$_3$ developing a materials' specific multiband, full bandwidth, anisotropic Eliashberg theory for this system.  
Our selfconsistent calculations highlight the importance of the interfacial electron-phonon interaction, which is hidden behind the seemingly weak coupling constant $\lambda_m$=0.4, in mediating the high-T$_{\rm c}$, and explain other puzzling experimental observations like the s-wave symmetry and replica bands. We discover that the formation of replica bands has a T$_{\rm c}$ decreasing effect that is nevertheless compensated by deep Fermi-sea Cooper pairing which has a T$_{\rm c}$ enhancing effect. We predict a strong coupling dip-hump signature in the tunneling spectra due to the interfacial coupling.
\end{abstract}

\maketitle

Superconductivity in monolayer-thick FeSe on SrTiO$_3$ reaches amazingly high transition temperatures of typically T$_{\rm c}$=50--70 K \cite{Qing-Yan2012,Liu2012,He2013,Tan2013,Peng2014a,Lee2014} and up to 100 K \cite{Ge2015}, much higher than the 8-K value of bulk FeSe \cite{Hsu2008}. 
A coupling between SrTiO$_3$ phonons and FeSe electrons occurs at the FeSe/SrTiO$_3$ interface, which manifests itself as electron replica bands \cite{Lee2014}. The value of this coupling is estimated by experiments to be around 0.4, thus it is commonly believed to moderately enhance T$_{\rm c}$ but not be enough to explain it \cite{Lee2014,Ding2016,Tian2016}.

The superconducting state in iron-based superconductors is customarily associated  with residual spin fluctuations due to the remnant quasi-nesting between electron and hole Fermi sheets that give rise to a sign alternating gap \cite{Mazin2008}. 
However, for FeSe/STO the situation is markedly different. 
Charge transfer at the interface induces electron doping in FeSe \cite{He2013,Tan2013}, leading to a distinct Fermi surface consisting of only two electron sheets around the corners of the tetragonal Brillouin zone (M point) \cite{Li2014}. 
The observed anisotropic superconducting gap has a more conventional form with plain s-wave symmetry \cite{Fan2015} and is thus nodeless in the entire Brillouin zone.
Furthermore, angular resolved photoemission spectroscopy (ARPES) measurements \cite{Lee2014} reveal an interface-induced electron-phonon interaction (EPI) between FeSe electrons and polar STO phonons that is strongly peaked at the ${\bf q}$=0 phonon wavevector \cite{Lee2014,Aperis2011,Weger1999,Varelogiannis1998,Kulifmmodecuteclseci1994,Abrikosov1994}.
There is growing experimental evidence for the pivotal role of such interfacial phonons in 
engineering high-T$_{\rm c}$ heterostructures that involve FeSe \cite{Peng2014a,Ding2016,Rebec2017} or even FeAs \cite{Choi2017} monolayers.

Although \textit{ab initio} calculations confirm the existence of small-q phonons as a strictly interfacial phenomenon in FeSe/STO \cite{Li2014,Xie2015,Wang2016b,Zhou2016a} and indicate the importance of the coupling between substrate phonons and FeSe electrons \cite{Huang2017}, the estimated low value of the electron-phonon coupling constant ($\lambda \leq 0.4$) has been widely considered insufficient to explain the impressive T$_{\rm c}$ enhancement 
unless another, dominant pairing mechanism is at play \cite{Lee2014}. On the other hand, Eliashberg calculations within a single band model suggest that interfacial phonons may lead to the high T$_{\rm c}$ with a coupling of merely half of that estimated by experiments \cite{Rademaker2016}.
However, a materials' specific  theory of superconductivity
that can account for the interplay between multiple bands, doping and small-q phonons has not yet been developed.
It remains therefore unsolved to what extend and how such phonons contribute to the peculiar superconductivity in FeSe/STO.

\begin{figure*}[ht!]
\includegraphics[width=0.9\textwidth]{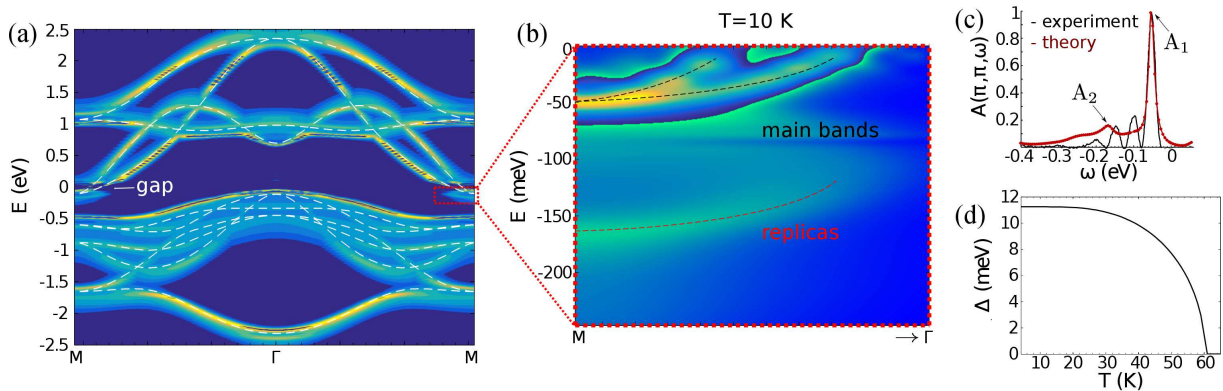}
\caption{Results of selfconsistent multiband Eliashberg theory calculations.
(a) Calculated full bandwidth spectral function at 10\,K, relevant to ARPES measurements, plotted along the M$-\Gamma -$M symmetry line of the Brillouin zone (BZ). The bare electron dispersions, used as input, are depicted by the white dashed lines. Several shake-off effects are visible in the quasiparticle band structure which are caused by interfacial EPI.
(b) Spectral function for small energies near the M point [=($\pi$,$\pi$)] of the BZ (red rectangle in a). Black dotted lines are guides to the eye. The formation of replica bands $110$ meV deeper than the main electron bands is clearly resolved \cite{Lee2014}. (c) The spectral function (red line) at M  exhibits a main peak at $-50$ meV and an additional, weaker peak near $-160$ meV.
The former peak is due to the main electron bands that form the Fermi surface in the normal state while the latter is their replica band contribution. The integrated experimental intensity at the M point
\cite{Lee2014} is shown by the black line. (d) Calculated temperature dependence of the superconducting gap edge maxima, giving $\Delta/ k_{\rm B} {\rm T_c} \approx 2.1$. 
}
\label{fig1}
\end{figure*}

Here, we present the first anisotropic, full bandwidth multiband Eliashberg calculations dedicated to unveil the influence of the interfacial electron-phonon coupling in FeSe/STO.
 Our theory extends on previous single band approaches \cite{Rademaker2016,Wang2016c} by establishing a microscopic description of 
superconductivity in this system on a materials' specific level, thus paving the way toward more realistic calculations of higher accuracy. Our selfconsistent results provide unambiguous support for the dominant contribution of these phonons to the high T$_{\rm c}$ and to further enigmatic experimental observations, and allows us to shed light on novel aspects of the mechanism responsible for the high-T$_{\rm c}$. Remarkably, bands not crossing the Fermi level provide an additional Cooper pairing channel that enhances T$_{\rm c}$ and places the value of the gap over T$_{\rm c}$ ratio in the strong coupling regime. 
In stark contrast to previous proposals, our here predicted deep Fermi sea Cooper pairing does not depend upon interband scattering processes mediated by bosons at large wavevectors, like e.g.\ antiferromagnetic spin fluctuations \cite{Bang2014,Linscheid2016a} or the conventional short-ranged in real space EPI \cite{Chen2015}.

In the presence of a strong inhomogeneous dielectric background, the EPI develops a pronounced forward scattering peak \cite{Abrikosov1994,Weger1996}. Here, the interface-induced EPI is modelled by a dispersionless mode at $\hbar \Omega$=81 meV \cite{Xie2015} coupled to FeSe electrons via the functional form $g({\bf q})=g_0 \exp{(-|{\bf q}|/q_c)}$ with $q_c=0.3a^{-1}$ \cite{Lee2014}, with $a$ the FeSe lattice constant.
 This is the \textit{only} mediator of superconductivity in our theory. In what follows, we do not take into account explicitly the effect of Coulomb repulsion on superconductivity, the implications of which are discussed further below. For the electron dispersions of monolayer FeSe we use a recently derived ten-band tight-binding bandstructure \cite{Hao2014,SM1}. Since the  FeSe doping level is not \textit{a priori} known, the electron filling is chosen such that the bottom of the electron bands around the M point in the Brillouin zone are at $-50$ meV as observed in experiment \cite{Liu2012,He2013,Lee2014}. 
 We determine the value of the electron-phonon scattering strength $g_0$, by requiring that the ARPES replica bands are reproduced at their observed energies \cite{Lee2014}. In this way, we circumvent the need for treating screening effects at the FeSe/STO interface explicitly \cite{Zhou2016a,Gorkov2016} and our determined value for the EPI strength may considered as the overall strength of the resulting \textit{effective} EPI. We obtain $g_0=728$ meV, which is significantly close to the \textit{ab initio} calculated value for anatase TiO$_2$ \cite{Verdi2015}, but a bit lower. This discrepancy may be understood as due to an enhanced screening effect at the FeSe/STO interface \cite{Zhou2016a}. We solve the three coupled Migdal-Eliashberg equations for  $\Delta({\bf k}, \omega)$, $Z({\bf k},\omega)$, and $\chi({\bf k},\omega)$, describing the superconductivity order parameter, electron mass and chemical potential renormalizations, respectively, selfconsistently with full bandwidth, momentum and energy dependence, while taking care to keep the electron occupancy $n$ fixed throughout the calculations \cite{SM1}. The latter quantity measures the electron filling (the case of half-filling corresponds to $n$=1). We find  $n$ $\approx$0.8, indicating that the system is in the electron-doped regime. We also note that when forward-scattering processes dominate the EPI, the Migdal theorem holds even in non-adiabatic cases \cite{Abrikosov2005}.
\begin{figure}[ht!]
\includegraphics[width=0.35\textwidth]{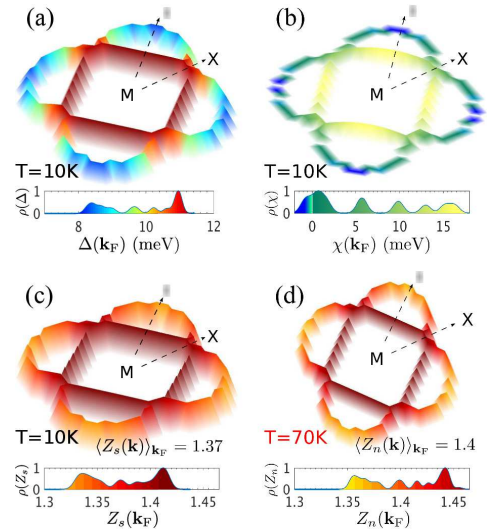}
\caption{Calculated momentum dependence of the superconducting gap and renormalization functions. 
(a) The gap edge $\Delta ({\bf k})$ is anisotropic ($\sim$25\%) over the Fermi surface with average value $\sim$10 meV. (b) The renormalized chemical potential $\chi ({\bf k})$ exhibits sign changes across the Fermi surface. 
(c) The mass renormalization function $Z ({\bf k})$ at T$<$T$_{\rm c}$ and (d) the same quantity at T$>$T$_{\rm c}$. 
{The insets show the distributions of the respective quantities over the Fermi surface.}
}
\label{fig2}
\end{figure}
\begin{figure*}[t!]
\includegraphics[width=0.8\textwidth]{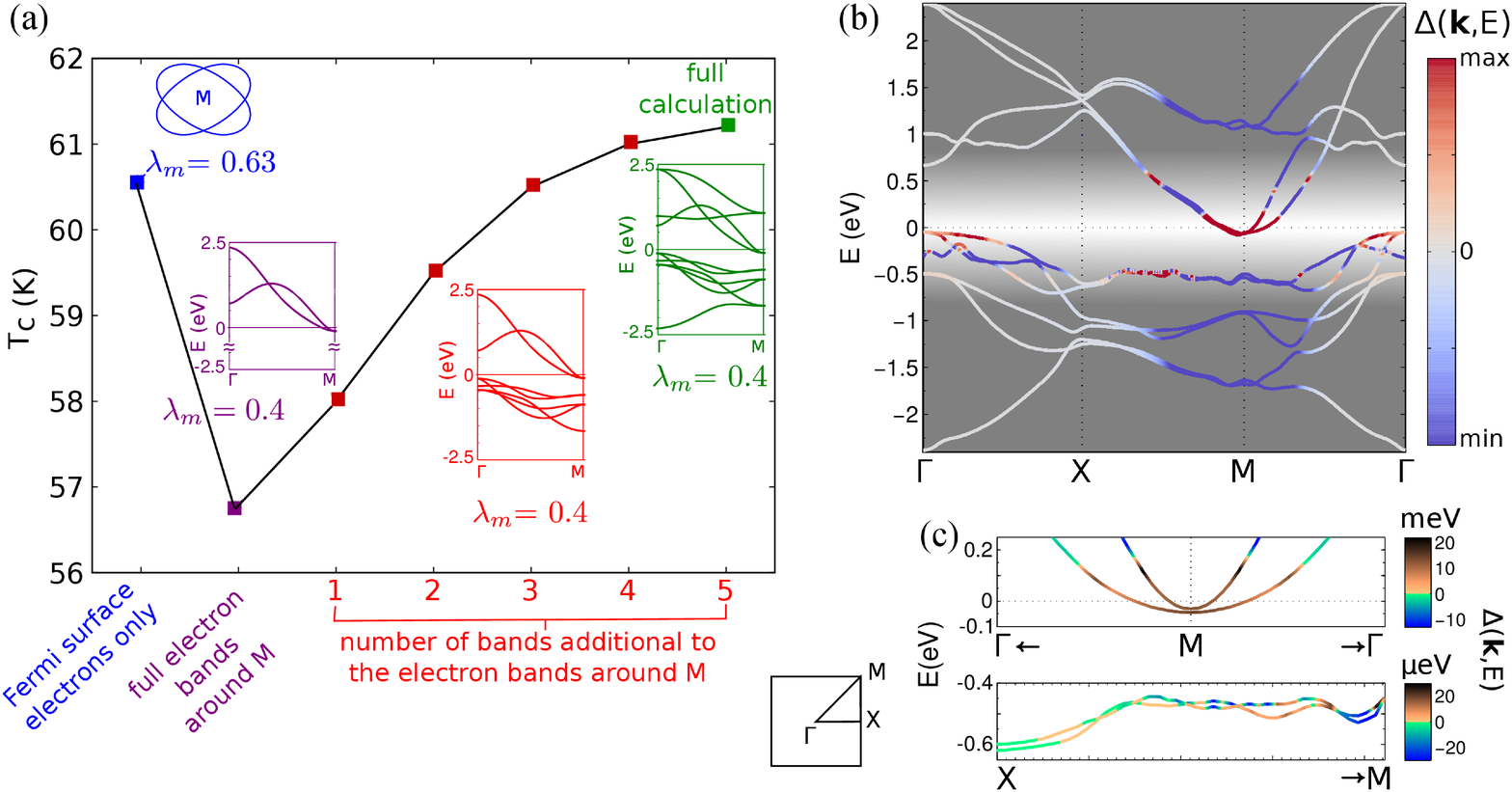}
\caption{Influence of included band dispersions on the superconductivity in monolayer FeSe/STO. (a) The calculated T$_{\rm c}$, starting from electrons restricted only to the Fermi surface (blue) and subsequently including their full bandwidth (purple) and electron and hole (red) band contributions until the full ten band calculation is recovered (green). Spectral weight transfer to the replica bands weakens $\lambda_m$ and therefore reduces T$_{\rm c}$ (purple). The interfacial phonon mode also couples weakly to bands away from the Fermi level, thus providing an additional channel to compensate this loss and even increase T$_{\rm c}$ (red and green). (b),(c) Cooper pair binding energy, $\Delta({\bf k},E)$, projected on each band of the renormalized electronic band structure due to the interfacial EPI. In (b) red (blue) color depicts electron attraction (repulsion) and thus Cooper pair formation (breaking). Apart from the dominant contribution of electrons in bands forming the Fermi surface (M point), deep Fermi-sea pairing takes place near the $\Gamma$ point and along the X--M high symmetry line of the folded Brillouin zone. 
}
\label{fig3}
\end{figure*} 

Figure \ref{fig1}(a) shows the calculated electron spectral function at T=10 K for the whole energy bandwidth and momenta along the M$-\Gamma -$M high-symmetry line of the folded Brillouin zone,
which is measured in ARPES experiments. Comparison of the spectral function with the bare tight-binding band structure  \cite{Hao2014} used as input in our calculations (shown by white dashed lines) reveals that the interfacial small-q phonon modifies the FeSe electrons in a manifest way.
Several shake-off effects in the band structure take place over the whole bandwidth, including the appearance of new bands near $-160$ meV. The opening of a superconducting gap around the Fermi level can be seen.
Most of the predicted shake-off effects in the band structure have not yet been observed by experiment \cite{Wang2016c}. However, the replica bands appearing near the M point (zoom-in shown in Fig.\ \ref{fig1}(b) at $\sim$110 meV distance from the main electron bands that form the Fermi surface have been experimentally resolved \cite{Lee2014}. Figure \ref{fig1}(c) highlights that not only the position of the peaks but also the peak ratio, $\rm A_2/A_1$=0.17, agrees well with experiment \cite{Lee2014}.
Notably, within solely phononic small-q theory we obtain the superconducting T$_{\rm c}$=61 K, in good agreement with experiment  \cite{Lee2014,Liu2012}, with a temperature dependence of the gap edge as shown in Fig.\ \ref{fig1}(d).  

The momentum dependence of the calculated superconducting gap $\Delta ({\bf k}_{\rm F})$ is shown in Fig.\ \ref{fig2}(a). It has s-wave symmetry and is moderately anisotropic ($\sim$25\%) with gap values that vary from $8-11$ meV over the Fermi surface with an average value of 10 meV. These values are in agreement with experiments \cite{Lee2014,Liu2012} although the location of the gap maxima seems to somewhat deviate from those experimentally measured (e.g.\ \cite{Lee2014}). The resulting anisotropy of the gap is a consequence of the small-q form of the interfacial EPI (cf.\ \cite{Varelogiannis1996}). Taking the maximum required excitation energy at the gap edge to calculate the gap over T$_{\rm c}$ ratio, we obtain the strong coupling (non-BCS) value $\Delta/k_{\rm B}{\rm T_{\rm c}}$=2.1 (in contrast to the BCS value $\Delta / k_{\rm B}{\rm T_{\rm c}}$=1.76).
The chemical potential renormalization $\chi({\bf k}_{\rm F})$, shown in Fig.\ \ref{fig2}(b), has an anisotropic momentum dependence with an average Fermi surface value of $\langle\chi({\bf k}_{\rm F})\rangle$=5.9 meV. The fact that $\chi({\bf k}_{\rm F})$ even changes sign at certain Fermi surface points indicates the highly non-trivial role this quantity plays in shaping the quasiparticle band structure of the monolayer. In contrast, the  mass renormalization function $Z({\bf k}_{\rm F})$, shown in Figs.\ \ref{fig2}(c),\,(d), is rather isotropic with an average Fermi surface value $\langle Z({\bf k}_{\rm F})\rangle$=1.37 and 1.40 for temperature below and above T$_{\rm c}$, respectively. This quantity is related to the electron-phonon coupling constant $\lambda_m$, by $\langle Z ({\bf k}) \rangle _{{\bf k}_{\rm F}} |_{\rm T>T_c}$=$1+\lambda_m$ which in our case yields $\lambda_m$= 0.4. This weak coupling value matches remarkably well to experiments \cite{Lee2014,Tian2016}.
 Also, in the superconducting state $\lambda^{\rm 10 K}_m\approx$0.37 
and satisfies $\rm A_2/A_1\approx\lambda^{\rm 10 K}_m/2$ (Fig.\ \ref{fig1}(c))  \cite{Kulic2017}.

 A plain calculation of T$_{\rm c}$ with our obtained value of $\lambda_m$=0.4 in McMillan's formula gives T$_{\rm c}$=17 K. On the other hand, using our numerical results in the two T$_{\rm c}$ formulas recently proposed for interfacial phonon-mediated superconductivity in FeSe/STO \cite{Kulic2017,Rademaker2016}, yields T$_{\rm c}$=272--283 K and 117.5 K, respectively (note that the T$_{\rm c}$ equation in \cite{Rademaker2016} is derived in the $q_c\rightarrow 0$ limit). These estimations are in stark contrast to the T$_{\rm c}$=61 K obtained here by our selfconsistent Eliashberg theory, which thus  resolves the controversy between a seemingly weak $\lambda_m$  and high-T$_{\rm c}$ superconductivity in FeSe/STO.
\begin{figure}[t!]
\includegraphics[width=0.44\textwidth]{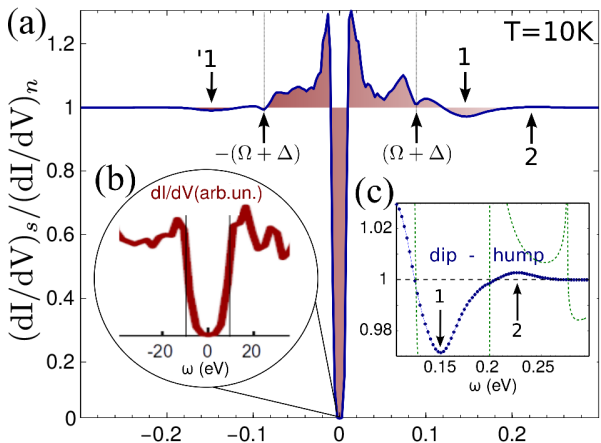}
\caption{Calculated tunneling spectra and predicted dip-hump signatures.
(a) The differential conductance dI/dV in the energy regime relevant to superconductivity normalized by the normal state value. A full gap with main coherence peaks at $\Delta$\,$\approx$\,$\pm$11 meV is in agreement with STS observations of plain s-wave superconductivity \cite{Fan2015}. The kinks near $\pm(\Delta+\Omega)$\,$\approx$\,$\pm$91 meV are signatures of the underlying electron-boson interaction, caused by the interfacial phonon mode. Additional features with the form of a dip-hump structure appear at higher energies, marked with 1('1) and 2, respectively. (b) The dI/dV in the low energy regime for direct comparison with STS data \cite{Fan2015}. (c) A zoom-in of the dip-hump region of the normalized dI/dV reveals a dip with minima near 150 meV and a hump with maxima near 220 meV. Within standard isotropic Eliashberg theory (green dotted line), this dip-hump signature corresponds to a ratio $\Delta /k_{\rm B}{\rm T_c}$\,$\approx$\,2.1 but with a strong coupling value of $\lambda_{iso}$=1.6.}
\label{fig4}
\end{figure}

To elucidate further the mechanism of T$_{\rm c}$ enhancement in FeSe/STO, we carried out a series of simulations where we first solve the usual momentum-dependent Eliashberg equations for electrons only at the Fermi surface of FeSe/STO and then perform full-bandwidth calculations while sequentially adding more bands until we recover the full bandwidth multiband calculation. Our findings are summarized in Fig.\ \ref{fig3}(a). In the first case (blue symbol in Fig.\ \ref{fig3}(a) where no electronic spectral rearrangement is allowed and thus no replica bands can form, we find $\lambda_m$=0.63. We note that this value equals the one given by the standard formula, $\lambda=\langle\lambda_{\bf q}\rangle_{{\bf k}_{\rm F}^{ },{\bf k}^{\prime}_{\rm F}}$=0.63 (with $\lambda_{\bf q}$  the momentum-dependent electron-phonon coupling \cite{SM1}). Therefore, the obtained T$_{\rm c}$=60.6 K is the maximum T$_{\rm c}$ reachable by Fermi-surface Cooper pairing.
Inclusion of the complete contribution of the two bands that form the Fermi surface (purple square in Fig.\ \ref{fig3}(a) leads to $\lambda_m$=0.4 and T$_{\rm c}$=56.8 K. Compared with the previous case,  here a part of the electron-phonon coupling strength is consumed in mediating the electronic spectral weight transfer from the bands crossing the Fermi level to the replica bands. The effective interaction left available for Cooper-pair mediation is weaker and concomitantly so is the T$_{\rm c}$. This weak coupling picture is further witnessed by the near-BCS value of the calculated ratio $\Delta/k_{\rm B}{\rm T_c}$=1.8. Remarkably, turning on contributions from near-Fermi-level bands (red symbols in Fig.\ \ref{fig3}(a) gradually increases T$_{\rm c}$ to 61.2 K (green square) but without affecting the value of $\lambda_m$. This behavior indicates that these bands contribute to superconductivity.

To quantify this remarkable finding, we projected the superconducting gap function $\Delta({\bf k},E)$ on the different electronic bands of FeSe/STO as shown in Fig.\ \ref{fig3}(b). Since $\Delta({\bf k},E)$ is a measure of the Cooper-pair binding energy,  it is positive for Cooper pairing with an s-wave gap  and negative otherwise \cite{Scalapino1966}. Figure \ref{fig3}(b) clearly shows that superconductivity in FeSe/STO stems not only from Fermi surface regions around M, but also from regions around the $\Gamma$-X and X-M directions of the Brillouin zone. Although in the latter regions $\Delta({\bf k},E)$ is in the $\mu$eV range (Fig.\ \ref{fig3}(c)), the resulting net contribution is enough to overcompensate the T$_{\rm c}$ decreasing effect of the replica band formation. It is also enough to raise the $\Delta/k_{\rm B}{\rm T_c}$ ratio to 2.1 and thus provide a strong coupling phenomenology.

We emphasize that our predicted deep Fermi sea Cooper pairing is markedly different from previous suggestions of pairing through incipient bands \cite{Bang2014,Chen2015,Linscheid2016a}. Here, the mediating interaction is not only phonon-driven but more importantly, it is local in momentum space due to its small-q shape, thus it relies explicitly on intraband processes.
 However, by the nature of our full bandwidth Eliashberg theory, different bands are coupled \textit{implicitly} via the frequency sector by the interfacial EPI, due to the large characteristic energy scale of the latter, in some sense reminiscent of the incipient band scenario.
  Our findings thus generalize the usual picture where Cooper pairing relies on near Fermi surface electrons and prove that deep Fermi-sea Cooper pairing is possible in multiband systems \cite{Bang2014,Chen2015,Linscheid2016a}. The recent puzzling superconductivity observed in doped LiFeAs without a Fermi surface  \cite{Miao2015} is plausibly explained within our picture.

Scanning tunneling spectroscopy (STS) measurements represent another key experimental feature reported for FeSe/STO \cite{Qing-Yan2012,Fan2015,Tang2016}.  To compare to STS data we have calculated the differential conductance spectrum $\rm dI/dV$ which, at low temperatures, is proportional to the superconducting density of states. 
The tunneling spectrum, calculated at T=10 K, is shown in Fig.\ \ref{fig4}.
 Zooming-in to the low energy regime (Fig.\ \ref{fig4}(b)) reveals the opening of an s-wave superconducting gap in the tunneling spectra that starts to close already around $\pm$5 meV and exhibits main coherence peaks at $\pm$11 meV, with secondary peaks a few meV's higher.
  The calculated spectrum is in excellent agreement with STS measurements \cite{Qing-Yan2012,Fan2015,Tang2016}. The position of the main coherence peaks in Fig.\ \ref{fig4}(b) coincides with the maximum of the gap-edge on the Fermi surface whereas the closing of the gap beginning at 5 meV is consistent with the minimum gap-edge of 8 meV (see Fig.\ \ref{fig2}(a)) in combination with the finite broadening of $\sim$3 meV used in our calculations.

Figure \ref{fig4}(a) shows the calculated tunneling spectra, normalized to the normal state values, at an intermediate energy range.
Remarkably, we find superconductivity related structures in the spectrum up to energies almost 30 times higher than the superconducting gap itself. Although it is well established that such non-BCS behavior is the hallmark of strong-coupling superconductivity \cite{Scalapino1966}, this is unexpected here given the seemingly weak coupling constant of the system.
In strongly coupled superconductors, the structure of the spectral function of the mediating bosons can be visible in the tunneling spectrum \cite{Scalapino1966}. Here, we predict that the interfacial phonon mode should manifest itself as two kinks around $\pm$91 meV (Fig.\ \ref{fig4}(a)), whose location coincides with $\pm(\Omega+\Delta)$ where $\Delta$ is the average gap-edge value. Furthermore, at higher energies we predict a distinct dip-hump structure in the spectra with a dip at 150 meV and a hump at 220 meV (Figs.\ \ref{fig4}(a),(c)).
Analyzing this additional strong coupling feature we find that it originates from the competition between the real and imaginary components of the superconducting gap function at an energy scale that is larger than the characteristic boson frequencies \cite{Varelogiannis1995}.
The energy location of the dip and the hump depends on the coupling strength \cite{Varelogiannis1995}, and more specifically, on the $\Delta/k_{\rm B}{\rm T_c}$ ratio \cite{SM1}.
For comparison, within \textit{isotropic} Eliashberg theory assuming an Einstein phonon at $\hbar \Omega$=81 meV, 
we find that our predicted dip-hump spectrum in FeSe/STO can only be fitted when the obtained gap over T$_{\rm c}$ ratio matches the one in FeSe/STO $\Delta/k_{\rm B}{\rm T_c}$=2.1 but with a strong-coupling isotropic value $\lambda _{iso}$=1.6 (Fig.\ \ref{fig4}(c)).

 The very good quantitative agreement between experiment and our selfconsistent calculations for FeSe/STO provides a consistent picture where the interfacial phonons drive the superconductivity.
 For that picture to be complete one needs to incorporate the pair-breaking effect of the Coulomb interaction on the T$_{\rm c}$. Inclusion of the latter effect into the full bandwidth Eliashberg calculations on an equal footing with the EPI requires knowledge of the frequency dependent renormalization of the Coulomb interaction throughout the system's bandwidth and is out of the scope of the present work. By approximating the Coulomb repulsion through the pseudopotential term $\mu^*$ \cite{SM1}, we estimate that for $\mu^*$= 0.1 and 0.14, T$_{\rm c}$=33 K and 26 K, respectively. However, the presence of an additional low-energy attractive channel due to the intrinsic EPI in FeSe monolayer \cite{Subedi2008,Koufos2014,1367-2630-17-7-073027}, although not sufficient to mediate the high-T$_{\rm c}$ on its own, can balance the T$_{\rm c}$ decrease due to Coulomb repulsion. We find that inclusion of the intrinsic EPI of freestanding monolayer FeSe \cite{1367-2630-17-7-073027} leads to T$_{\rm c}$=57 K and 51 K for $\mu^*$= 0.1 and 0.14, respectively.
 
In conclusion, our first of its kind full-bandwidth multiband theory shows that the interfacial EPI in FeSe/STO with a seemingly weak $\lambda_m$=0.4, explains key experimental facts like the replica bands, superconducting gap and tunneling spectra while also producing the correct T$_{\rm c}$ in the absence of any significant Coulomb pair-breaking. Our explicit calculations unveil the T$_{\rm c}$ increasing effect of deep Fermi-sea Cooper pairing and the T$_{\rm c}$ decreasing effect of replica-band formation,
and suggest new pathways to engineer high T$_{\rm c}$'s. A definite confirmation for the former effect will be the observation of a dip-hump feature in the tunneling spectra, which will also serve as an additional fingerprint of the decisive involvement of the interfacial EPI in mediating the high-T$_{\rm c}$.
On a fundamental level, our findings put to the question the current standard perception of the
efficiency of EPI in mediating high-T$_{\rm c}$ superconductivity and, whether Fermi-surface restricted theory is sufficient to capture the superconductivity of other doped high-T$_{\rm c}$ materials.

We thank G.\ Varelogiannis for fruitful discussions. This work has been supported by the Swedish Research Council (VR), the R{\"o}ntgen-{\AA}ngstr{\"o}m Cluster, and the Swedish National Infrastructure for Computing (SNIC).

\bibliographystyle{apsrev4-1}

\end{document}